 \pdfoutput=1
\documentclass[%
preprint,
 amsmath,amssymb,
 aps,
floatfix,
]{revtex4-1}
\pdfoutput=1
\usepackage{appendix}
\usepackage{amsmath}
\usepackage{graphicx}
\usepackage{epstopdf}
\usepackage{graphics}
\usepackage{epsfig}
\usepackage[byname]{smartref}
\usepackage{hyperref} %comment out for hardcopy
\usepackage{txfonts}
\usepackage{morefloats}
\usepackage{tocloft}
\usepackage{graphicx}% Include figure files
\usepackage{dcolumn}% Align table columns on decimal point
\usepackage{bm}% bold math
\usepackage{hyperref}% add hypertext capabilities
\begin{document}

\preprint{V1}

\title{Some Properties of Bilayer Graphene Nanoribbons}% Force line breaks with \\

\author{Maher Z. Ahmed}
\affiliation{The National Institute of Standard
\\Giza, Egypt}
%\affiliation{Physics Department, Faculty of Science, Ain Shams University, Abbsai, Cairo, Egypt}
\email{maher2100@gmail.com}
%
%\author{M. G. Cottam}
%\email{cottam@uwo.ca} \affiliation{Department of Physics and Astronomy, University of Western Ontario, London ON N6A 3K7, Canada}

%\author{R. N. Costa Filho}
%\email{rai@fisica.ufc.br}
%\affiliation{Department of Physics and Astronomy, University of Western Ontario, London ON N6A 3K7, Canada}
%\affiliation{Departamento de F\'{\i}sica, Universidade Federal do Cear\'{a}, Caixa Postal 6030, Campus do Pici, 60455-760 Fortaleza, Ceara, Brazil}

%\date{September 6, 2010}% It is always \today, today,

\begin{abstract}
In this work the tight binding model calculations are carried out for AA-Bilayer Graphene
nanoribbons as an example of bilayer systems. The effects of edges, NNN
hopping, and impurities of a single layer are introduced numerically as a
change in the elements of the relevant block diagonal matrix appearing in the
direct diagonalization method. The direct interlayer hopping between the top
and the bottom single layers is constructed in the generalized direct
diagonalization method by the off-diagonal block matrices in which the
strength of the interlayer hopping is included.
\end{abstract}

\pacs{Valid PACS appear here}% PACS, the Physics and Astronomy
\maketitle

%\tableofcontents

% ------------------------------------------------------------------------
% -*-TeX-*- -*-Hard-*- Smart Wrapping
% ------------------------------------------------------------------------
\def\baselinestretch{1}

\section{Introduction}
In the previous work \cite{Ahmed5,Ahmed4}, we have studied the effects of lattice structures
(including both the honeycomb and square lattices), the interaction range (NN
and NNN), and the presence of impurities on the dispersion relations of the
2D materials. This was carried out both for graphene nanoribbons and magnetic
stripes in order to compare and contrast their behavior. The intrinsic
physical properties of 2D materials may not, in general, be easily tunable
and therefore they cannot necessarily meet many technological applications
design requirements. However, it has been proposed in the literature that a
system of two graphene layers stacked on top of each other might give rise to
the possibility of controlling their physical properties by introducing
asymmetry between the layers. This could be done in various ways like varying
the external electric or magnetic field, rotation between the two layers, and
introducing impurities in one layer
\cite{Castro2010,PhysRevB.80.045308,kalon:233108,doi:10.1021/nl902948m,Ohta18082006}.
This opens the possibility of technological applications using bilayer
graphene \cite{Chakraborty}.

In this work, we will extend our formalism for single-layer graphene as
developed in \cite{Ahmed4,Ahmed5} to examine the effect of forming a system of
two layers stacked on top of each other. We then derive the dispersion
relations and study the localized edge modes. Specifically, the system that
will be used for our study consists of two graphene layers stacked directly
on top of each other to form what is called AA-stacking bilayer graphene
(BLG) nanoribbons. This system is interesting both experimentally and
theoretically
\cite{PhysRevLett.102.015501,Fang-Ping2011,Chang2010,PhysRevB.77.045403}.

The tight binding Hamiltonian \cite{Chakraborty,Neto1}  will be used as
before to describe the NN and NNN hopping in each graphene single layer
(GSL), while additional tight binding terms are introduced to describe the
direct hopping between the two layers \cite{Chakraborty,Neto1}. As well as
the main applications to graphene bilayer nanoribbons, the results should
also be capable of extension to similar magnetic stripes in bilayer
configurations which could be fabricated as "nanodot" arrays
\cite{Jiang1996,PhysRevB.77.180506,PhysRevB.71.184436,carapella:242504}.

\section{Theoretical model}
\begin{figure}[h]
\centering
\centering
  \begin{tabular}{cc}
\includegraphics[scale=0.27]{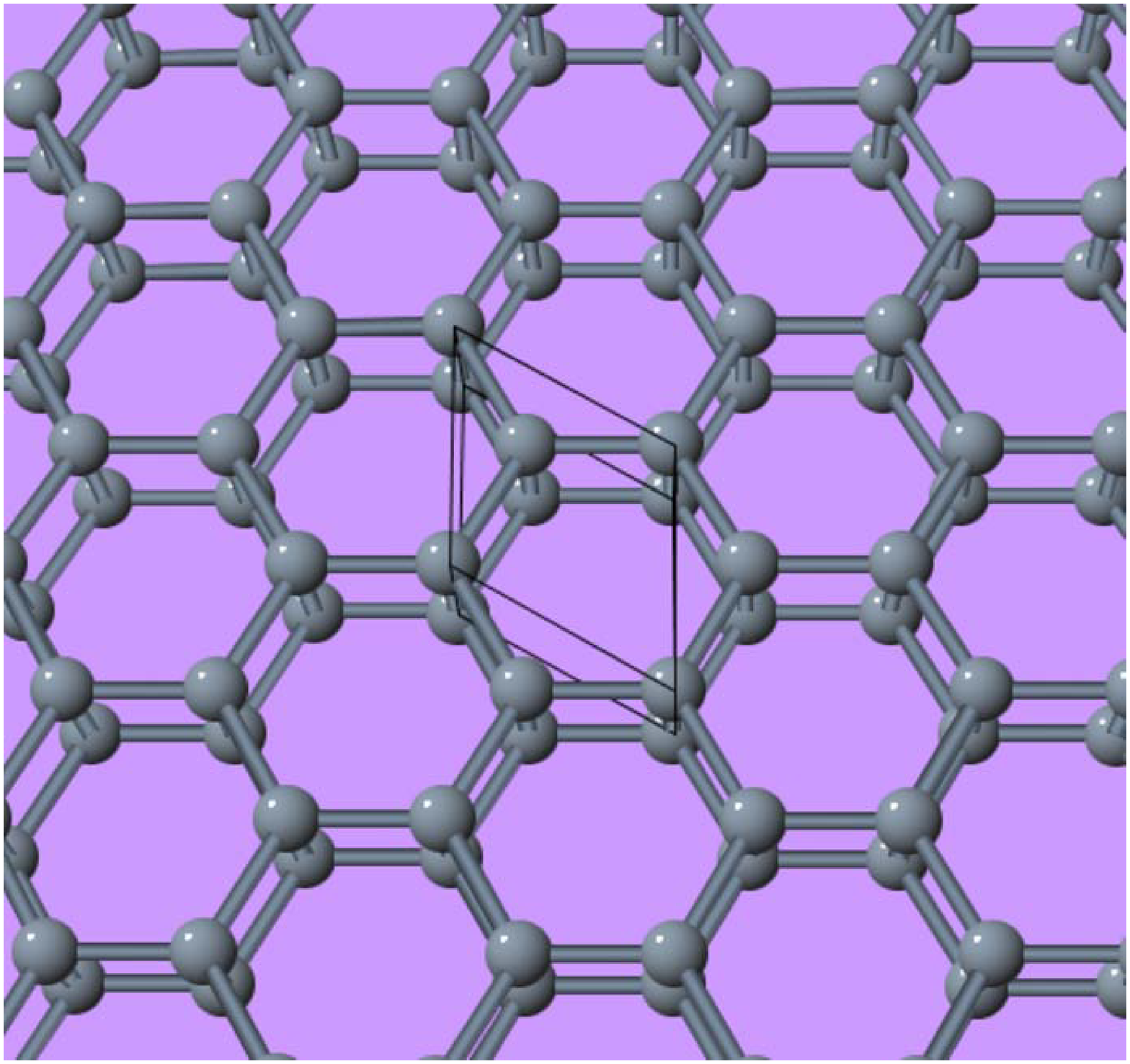}& \includegraphics[scale=0.8]{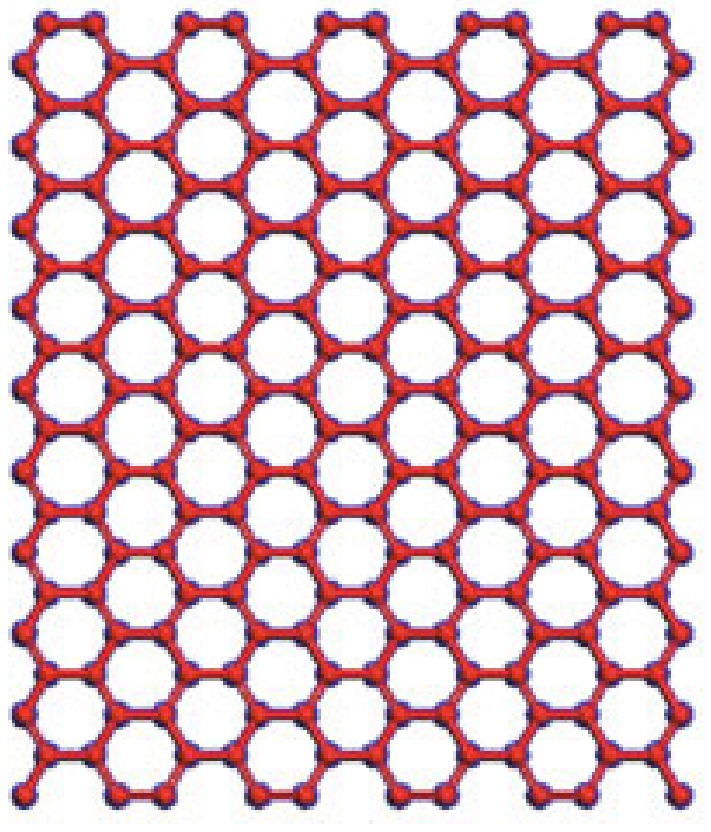}
\end{tabular}
\caption{Direct on-top AA-stacking bilayer graphene (BLG). Left: the 2D unit cell is shown
$a=b=0.267 nm$, $\gamma  =120^\circ$. Right: AA-stacking BLG nanoribbons.  Figures taken from \cite{PhysRevB.77.045403,Chang2010}.}\label{fig:graphenelattice}
\end{figure}

The system initially under study consists of two graphene layers stacked
directly on top of each other, i.e.,  AA-stacking bilayer graphene (BLG)
nanoribbons in the $xy$-plane, where we use the indices ``t" and ``b" to
label the top and bottom layer, respectively. The crystallographic
description of each graphene layer with its honeycomb lattice is given in
\cite{Chakraborty,Neto1}. The bilayer
nanoribbon is of finite width in the $y$ direction with $N$ atomic rows
(labeled as $n = 1,\cdots,N$) and it is infinite in the $x$ direction (see
Figure \ref{fig:graphenelattice}).

The total Hamiltonian of the system is given as follows:
\begin{equation}\label{secondhamiltonian137}
\hat{H}=\hat{H}_t+\hat{H}_b+\hat{H}_{i},
\end{equation}
where $\hat{H}_t$ ($\hat{H}_b$)  is the Hamiltonian of the top (bottom)
single layer of graphene (SLG) which describes the in-plane hopping of
non-interacting $\pi$-electrons on the top (bottom) layer and it is given by
\begin{eqnarray}
\nonumber\hat{H}_t&=&-\sum_{\langle ij,t\rangle} t_{{0}_{ij},t} (\textbf{a}^\dag_{i,t} \textbf{b}_{j,t}+\textbf{h.c})+t_{{1}_{ij},t}(\textbf{a}^\dag_{i,t} \textbf{a}_{j,t}+\textbf{b}^\dag_{i,t} \textbf{b}_{j,t}+\textbf{h.c}),\\
\hat{H}_b&=&-\sum_{\langle ij,b\rangle} t_{{0}_{ij},b} (\textbf{a}^\dag_{i,b} \textbf{b}_{j,b}+\textbf{h.c})+t_{{1}_{ij},b}(\textbf{a}^\dag_{i,b} \textbf{a}_{j,b}+\textbf{b}^\dag_{i,b} \textbf{b}_{j,b}+\textbf{h.c}).
\end{eqnarray}

The notation is defined as follows the first term in each
layer $t_{0_{ij}}(\approx 2.8 \textrm{eV})$ is the NN hopping energy, and here in graphene it is the hopping between
different sublattices $A$ and $B$. Also $t_{1_{ij}} (\approx 0.1
\textrm{eV})$ is the NNN hopping energy in each layer which here in graphene
is the hopping in the same sublattice \cite{Neto1,Deacon2007,Reich2002}. The
summations in each layer over $i$ and $j$ run over all the sites where $i$
and $j$ belong to different sublattices for the NN hopping term, and they
belong to the same sublattice for the NNN hopping energy. The labeling schemes for the various hopping terms (at an
edge or in the interior) are the same as in the previous work for the
single-layer case.

The third term in the total Hamiltonian $\hat{H}_{i}$ represents the
Hamiltonian of the direct inter-layer hopping between the top and the bottom
single layers where the sublattice $A_t$ ($B_t$) of the top layer is directly
above the sublattice $A_t$ ($B_t$) of the bottom layer. It is given by
\begin{equation}
  \hat{H}_{i}=  -\sum_{\langle i\rangle} \gamma_{0}  (\textbf{a}^\dag_{i,b}\textbf{a}_{i,t}+\textbf{b}^\dag_{j,b}\textbf{b}_{j,t}+\textbf{h.c}),
\end{equation}
where $\gamma_{0}$ the inter-layer NN coupling energy.

Since the nanoribbon extends to $\pm\infty$ in the $x$ direction, we may
introduce a 1D Fourier transform to wavevector $q_x$ along the $x$ direction
for the fermions operators $a^{\dag}_{i}$ ($a_{i}$) and $b^{\dag}_{j}$
($b_{j}$) in each layer as follows:
\begin{align}
b_{j}(x)&=\frac{1}{\sqrt{N_0}} \sum_{n} b_{n}(q_x) e^{-i \mathbf{q}_x\cdot \mathbf{r}_j},\hspace{20pt} b_{j}^\dag(x)=\frac{1}{\sqrt{N_0}} \sum_{n} b^\dag_{n}(q_x) e^{i \mathbf{q}_x\cdot \mathbf{r}_j},  \nonumber\\
a_{i}(x)&=\frac{1}{\sqrt{N_0}} \sum_{n} a_{n}(q_x) e^{-i \mathbf{q}_x\cdot \mathbf{r}_i},\hspace{20pt} a_{i}^\dag(x)=\frac{1}{\sqrt{N_0}} \sum_{n} a^\dag_{n}(q_x) e^{i \mathbf{q}_x\cdot \mathbf{r}_i}.\label{far337}
\end{align}
Here, $N_0$ is the (macroscopically large) number of carbon sites in any row,
$\mathbf{q_x}$ is a wavevector in the first Brillouin zone of the reciprocal
lattice and both $\mathbf{r}_i$ and $\mathbf{r}_j$ denote the position
vectors of any carbon sites $i$ and $j$. The new fermion operators obey the
following anticommutation relations in each layer:
\begin{equation}\label{comu337}
\left[a_{n}(q_x),a^\dag_{n}(q'_x)\right]=\delta_{q_xq'_x},\hspace{30pt}\left[b_{n}(q_x),b^\dag_{n}(q'_x)\right]=\delta_{q_xq'_x},
\end{equation}
while the top layer operators anticommute with the bottom layer operators.
Also, we define the hopping sums
\begin{eqnarray}
\nonumber
\tau(q_x) &=& \sum_{\nu} t_{{0}_{ij}}  e^{-i\mathbf{q}_x \cdot (\mathbf{r}_i-\mathbf{r}_j)},\\
\tau'(q_x) &=&\sum_{\nu'} t_{{1}_{ij}}  e^{-i\mathbf{q}_x \cdot (\mathbf{r}_i-\mathbf{r}_j)} \label{hoppingsum7}.
\end{eqnarray}
The sum for the hopping terms $t_{{0/1}_{ij}}$ is taken to be over all $\nu$
NNs and over all $\nu'$ NNNs in the lattice and this depends on the edge
configuration as zigzag or armchair for the stripe (see \cite{Ahmed5}). For the inter-layer coupling the
in-plane wavevector $\mathbf{q}_x$ is perpendicular to the inter-plane vector
and so the hopping sum is only $\gamma_{0}$ representing the corresponding NN
coupling energy.

For the armchair configuration, the hopping sum for NNs gives the following
factors $ \tau_{nn'}(q_x)$ for each (top or bottom) layer:
\begin{eqnarray}
  \tau_{nn'}(q_x)=t\left[\exp(iq_xa)\delta_{n',n}+\exp\left(i\frac{1}{2}q_xa\right)\delta_{n',n\pm1}\right]
\end{eqnarray}
and for the zigzag configuration, it gives:
\begin{eqnarray}
  \tau_{nn'}(q_x)=t\left[2\cos\left(\frac{\sqrt{3}}{2}q_xa\right)\delta_{n',n\pm1}+\delta_{n',n\mp1}\right].
\end{eqnarray}

Explicitly the hopping sum for NNNs gives the following factors
$\tau'_{nn'}(q_x)$ for each layer
\begin{eqnarray}
  \tau_{nn'}(q_x)=t'\left[\delta_{n',n\pm2}+ 2\cos(q_x a3/2)\delta_{n',n\pm1}\right]
\end{eqnarray}
for the armchair configuration, and
\begin{eqnarray}
  \tau_{nn'}(q_x)=2t'\left[\cos(\sqrt{3}q_x a)\delta_{n',n}+\cos(\sqrt{3}q_x a/2)\delta_{n',n\pm2}\right]
\end{eqnarray}
for the zigzag configuration case, where the $\pm$ sign, in all the above
factors, depends on the sublattice since the atom lines alternate between the
A and B sublattices.

Substituting Equations~\eqref{far337} and \eqref{hoppingsum7}  in
Equation~\eqref{secondhamiltonian137}, and rewriting the summation over NN
and NNN sites, we get the following form of the total Hamiltonian operator
$\hat{H}$:
\begin{eqnarray}% \nonumber
\hat{H}&=&-\sum_{nn',t}  \tau'(q_x)_t \left(a^\dag_{n,t} a_{n',t}+ b^\dag_{n,t} b_{n',t} \right) + \tau(q_x)_t a_{n,t}  b^\dag_{n',t}+ \tau(-q_x)_t a^\dag_{n,t}  b_{n',t}  \nonumber \\
       &&-\sum_{nn',b}  \tau'(q_x)_b \left(a^\dag_{n,b} a_{n',b}+ b^\dag_{n,b} b_{n',b} \right) + \tau(q_x)_b a_{n,b}  b^\dag_{n',b}+ \tau(-q_x)_b a^\dag_{n,b}  b_{n',b}   \nonumber \\
       && -\sum_{n} \gamma_{0}  (a^\dag_{n,b} a_{n,t} +  b^\dag_{n,b}b_{n,t}+H.c.),
\end{eqnarray}
where the first two lines account for the NN and NNN intra-layer hopping in
the t and b layers, respectively, and the third line represent inter-layer
hopping.

In order to diagonalize $\hat{H}$ and obtain the dispersion relations for
AA-stacking bilayer graphene (BLG) nanoribbons, we may consider the modified
time evolution of the creation and the annihilation operators $a^{\dag}_{i}$
($a_{i}$) and $b^{\dag}_{j}$ ($b_{j}$), as calculated in the Heisenberg
picture in quantum mechanics. In this case, the equations of motion (using
the units with $\hbar=1$) for the annihilation operators $a_{i}$($b_j$) are
as follows
\cite{Bes2007,Liboff,Kantorovich2004,Roessler2009,HenrikBruus2004}:
\begin{eqnarray}
  \frac{d a_{n,t}}{dt} &=&i[H,a_{n,t}]=i[H_t+H_b+H_{i},a_{n,t}] \nonumber \\
     &=&i\sum_{nn'}-\tau'(q_x)  a_{n',t}-\tau(-q_x) b_{n',t}-\gamma_{0}a_{n,b} \label{ch3em1371}
\end{eqnarray}
and
\begin{eqnarray}
  \frac{d b_{n,t}}{dt} &=&i[H,b_{n,t}]=i[H_t+H_b+H_{i},b_{n,t}] \nonumber \\
   &=&i\sum_{nn'}-\tau'(q_x)  b_{n',t}-\tau(q_x) a_{n',t}-\gamma_{0}b_{n,b}\label{ch3em1372}
\end{eqnarray}
for the operators of the top layer. Similarly for the bottom layer we have
\begin{eqnarray}
  \frac{d a_{n,b}}{dt}  &=& i\sum_{nn'}-\tau'(q_x)  a_{n',b}-\tau(-q_x) b_{n',b}-\gamma_{0}a_{n,t} \nonumber \\
  \frac{d b_{n,b}}{dt}  &=& i\sum_{nn'}-\tau'(q_x)  b_{n',b}-\tau(q_x) a_{n',b}-\gamma_{0}b_{n,t} \label{ch3em1373}
\end{eqnarray}
where Equation~\eqref{comu337} was used.

The electronic dispersion relations of the double-layer graphene  (i.e.,
energy or frequency versus wavevector) can now be obtained by solving the
above operator equations of motion. Assuming, as before, that the coupled
electronic modes behave like $\exp[-i\omega(q_x)t]$, on substituting this
time dependent form into Equations \ref{ch3em1371}-\ref{ch3em1373}, we get
the following set of coupled equations:
\begin{eqnarray}
\omega(q_x) a_{n,t}  &=&\sum_{n'}\tau'_{nn'}(q_x)  a_{n',t}+\tau_{nn'}(-q_x) b_{n',t}+\gamma_{0}a_{n,b} \nonumber \\
\omega(q_x) b_{n,t}  &=&\sum_{n'}\tau_{nn'}(q_x) a_{n',t}+\tau'_{nn'}(q_x)  b_{n',t} +\gamma_{0}b_{n,b} \nonumber \\
\omega(q_x) a_{n,b}  &=&\sum_{n'}\tau'_{nn'}(q_x)  a_{n',b}+\tau_{nn'}(-q_x) b_{n',b}+\gamma_{0}a_{n,t} \nonumber \\
\omega(q_x) b_{n,b}  &=&\sum_{n'}\tau_{nn'}(q_x) a_{n',b}+\tau'_{nn'}(q_x)  b_{n',b}+\gamma_{0}b_{n,t}
\end{eqnarray}
The above equations can be written in matrix form as follows:
\begin{eqnarray}
\omega(q_x)\left[%
\begin{array}{c}
a_{n,t}   \\
b_{n,t}   \\
a_{n,b}   \\
b_{n,b}   \\
\end{array}%
\right]  &=& \left[%
\begin{array}{cccc}
  T_t'(q_x)   & T_t(q_x)&U & 0 \\
  T_t^*(q_x) & T_t'(q_x)&0& U \\
U & 0 & T_b'(q_x)   & T_b(q_x)\\
 0 & U& T_b^*(q_x) & T_b'(q_x)\\
\end{array}%
\right] \left[%
\begin{array}{c}
a_{n,t}   \\
b_{n,t}   \\
a_{n,b}   \\
b_{n,b}   \\
\end{array}%
\right]\,\label{matrix234}
\end{eqnarray}
where the solution of this matrix equation is given by the determinantal
condition
\begin{eqnarray}
\det \left[%
\begin{array}{cccc}
-R(q_x)_t   & T_t(q_x)&U &0  \\
  T_t^*(q_x) & -R(q_x)_t&0& U \\
U & 0 & -R(q_x)_b   & T_b(q_x)\\
 0 & U& T_b^*(q_x) & -R(q_x)_b\\
\end{array}%
\right] =0 \label{matrix237}
\end{eqnarray}
where we denote $R(q_x)_t=\omega(q_x)I_N-T_t'(q_x)$ , and for each layer
$T(q_x)$ and $T'(q_x)$ are the NN and NNN interaction matrices respectively,
which depend on the orientation of the ribbon, and $\omega(q_x)$ are the
energies of the modes. The matrix $T(q_x)$ is given by
\begin{equation}\label{}
\left(
  \begin{array}{ccccc}
    \alpha & \beta   &     0 & 0 & \cdots \\
    \beta   & \alpha & \gamma& 0 & \cdots \\
    0       & \gamma  &\alpha&  \beta    &\cdots \\
    0       &      0  & \beta & \alpha & \cdots \\
     \vdots & \vdots& \vdots & \vdots & \ddots \\
  \end{array}
\right).
\end{equation}
whereas the NNN matrix $T'(q_x)$ for zigzag ribbons is given	by
\begin{equation}\label{}
\left(
  \begin{array}{cccccc}
    \epsilon& 0     &\zeta & 0    & 0 & \cdots \\
   0   &\epsilon & 0   & \zeta    & 0 & \cdots \\
    \zeta     &0 &\epsilon &  0 & \zeta &\cdots \\
    0     &  \zeta  & 0 & \epsilon& 0 & \cdots \\
    0     &   0 &   \zeta  & 0 & \epsilon & \cdots \\
     \vdots & \vdots& \vdots & \vdots & \vdots& \ddots \\
  \end{array}
\right)
\end{equation}
and the matrix $T'(q_x)$ for armchair ribbons is given by
\begin{equation}\label{}
\left(
  \begin{array}{cccccc}
    0& \eta    &\theta & 0    &0  & \cdots \\
   \eta   &0 & \eta  & \theta   &  0& \cdots \\
    \theta   &\eta &0 &  \eta & \theta &\cdots \\
    0     &  \theta& \eta & 0& \eta & \cdots \\
    0     &   0 &   \theta  & \eta& 0 & \cdots \\
     \vdots & \vdots& \vdots & \vdots & \vdots& \ddots \\
  \end{array}
\right)
\end{equation}

Finally, $U$ is the inter-layer hopping coupling matrix and it is given by
\begin{equation}\label{u}
\left(
  \begin{array}{cccccc}
    \gamma_{0}   & 0   &0 & 0    &0  & \cdots \\
    0   &\gamma_{0}    & 0  & 0  &  0& \cdots \\
    0   &0    &\gamma_{0} &  0 & 0 &\cdots \\
    0   &  0  & 0 & \gamma_{0}& 0 & \cdots \\
    0   &   0 &   0  & 0& \gamma_{0} & \cdots \\
     \vdots & \vdots& \vdots & \vdots & \vdots& \ddots \\
  \end{array}
\right)
\end{equation}
The parameters $\alpha,\beta,\gamma,\epsilon,\zeta,\theta$ and $\eta$ depend
on the stripe edge geometry and are given in Tables \ref{311} and \ref{312}.

In the special case that the NNN hopping $t'$ can be neglected compared to
the NN hopping $t$, the $T'(q_x)$ matrix is equal to the zero matrix
$\mathbf{0}$ and Equation~\eqref{matrix237} simplifies to become
\begin{eqnarray}
\det \left[%
\begin{array}{cccc}
-\omega(q_x)I_N   & T_t(q_x)&U &0  \\
  T_t^*(q_x) & -\omega(q_x)I_N&0& U \\
U & 0 & -\omega(q_x)I_N   & T_b(q_x)\\
 0 & U& T_b^*(q_x) & -\omega(q_x)I_N\\
\end{array}%
\right] =0
\end{eqnarray}

\begin{table}[h]
 \caption{NN hopping matrix elements for the graphene honeycomb
lattice}\label{311}
\begin{tabular}{lcl}
  \hline\hline
  Parameter& Zigzag               & Armchair \\ \hline
   $\alpha$&         0           &$te^{-iq_xa}$  \\
   $\beta$ &$2t \cos(\sqrt{3}q_x a/2)$&$te^{iq_xa/2}$\\
   $\gamma$& $t$ &$ te^{iq_xa/2}$ \\
  \hline
\end{tabular}
  \centering
 \caption{NNN hopping matrix elements for the graphene honeycomb
lattice}\label{312}
\begin{tabular}{lcll}
  \hline\hline
  Parameter& Zigzag  &  Parameter           & Armchair \\ \hline
  $\epsilon$& $2t' \cos(\sqrt{3}q_x a)$& $\theta$                  & $t'$ \\
  $\zeta$& $2t' \cos(\sqrt{3}q_x a/2)$ &$\eta$&$2t' \cos(q_x a3/2)$ \\
  \hline
\end{tabular}
  \centering
\end{table}

The dispersion relations for the above graphene nanoribbons are next obtained
numerically as the eigenvalues \cite{algebra,RefWorks:27} for the matrix
Equation \eqref{matrix234}. Again, this is formally similar to equations obtained in \cite{Ahmed2} and therefore the same numerical method used for analogous honeycomb magnetic stripes will be used here to get
the required solutions. The effects of edges and impurities in each layer can
be introduced into the numerical calculations just as in \cite{Ahmed4,Ahmed5}.

In the context of analogous bilayer magnetic materials, it is interesting to
note that the matrix Equation \eqref{matrix234} can also be modified to apply
to the 2D magnetic square lattice, taking into account the difference between
it and the honeycomb lattice, including the existence of only one type of
lattice site. Consequently the matrix size for the square lattice case is
reduced to give the condition
\begin{eqnarray}
\det\left[%
\begin{array}{cc}
-\omega(q_x)\alpha_t I_N+T_t'(q_x)+T_t(q_x)  & U   \\
 U & -\omega(q_x)\alpha_b I_N+T_b'(q_x)+T_b(q_x)
\end{array}%
\right] =0.
\end{eqnarray}

\section{Numerical results}

\begin{figure}[hp]
  \centering
  \begin{tabular}{cc}
\includegraphics[scale=.6]{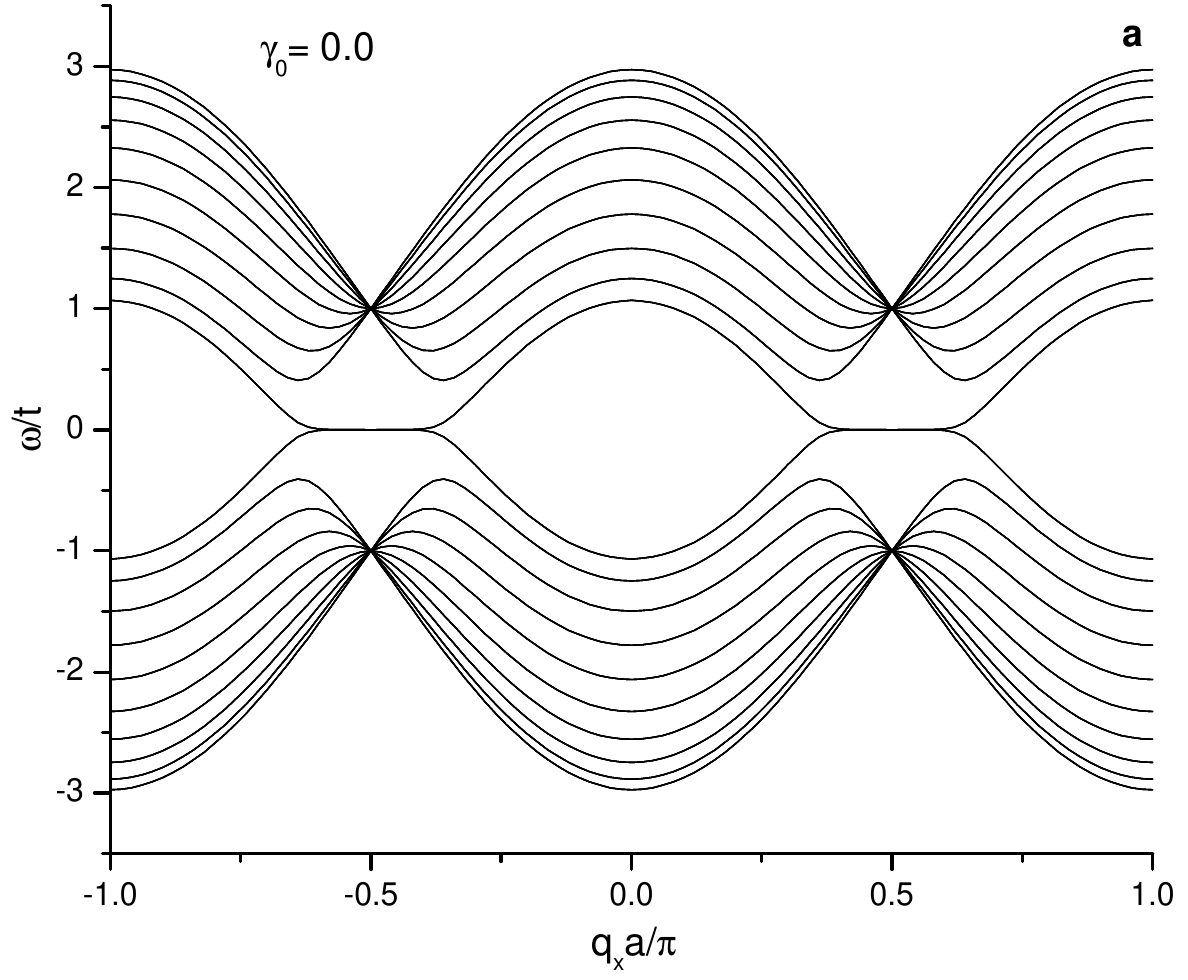}& \includegraphics[scale=.6]{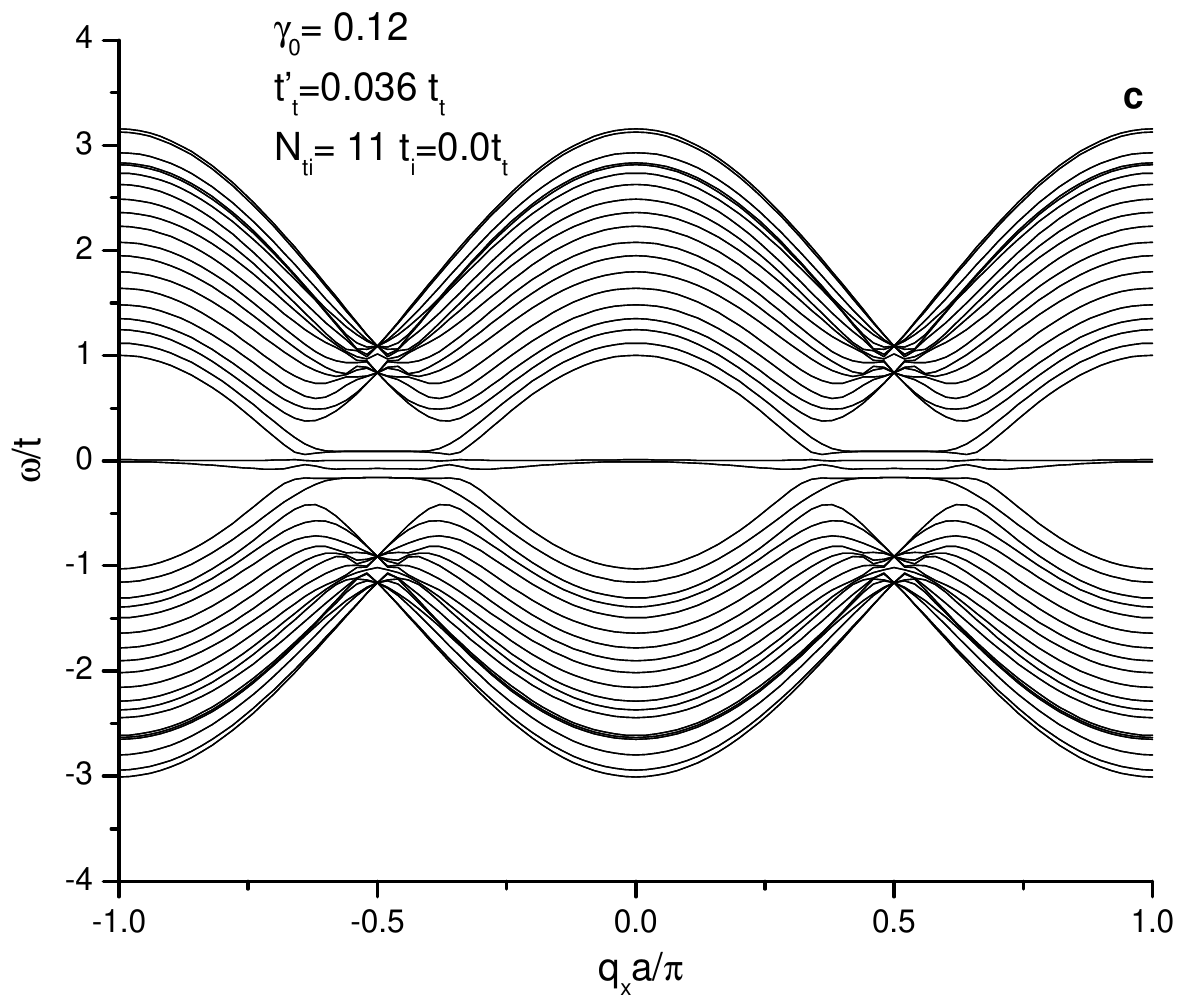}\\
\includegraphics[scale=.6]{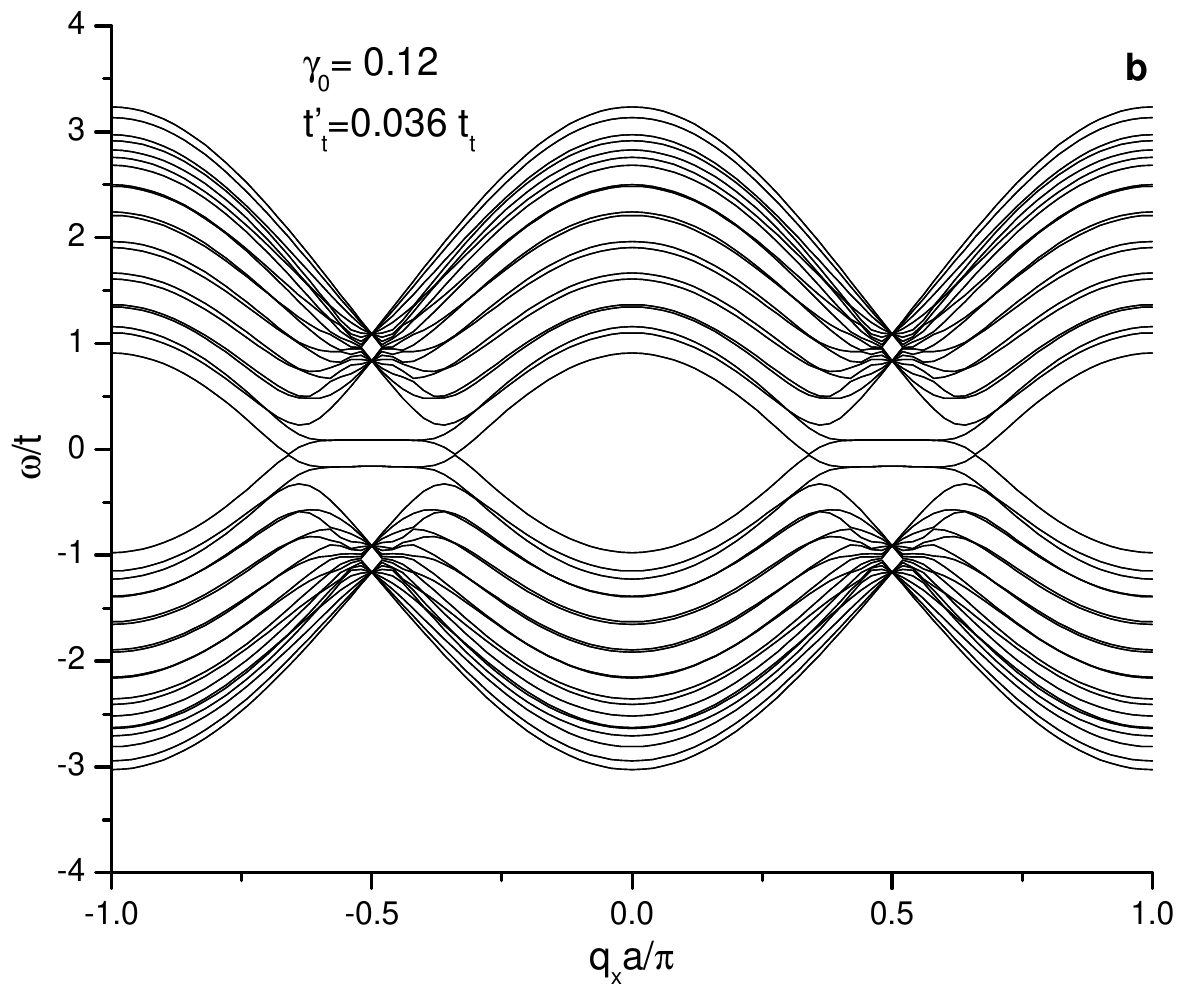}&\includegraphics[scale=.6]{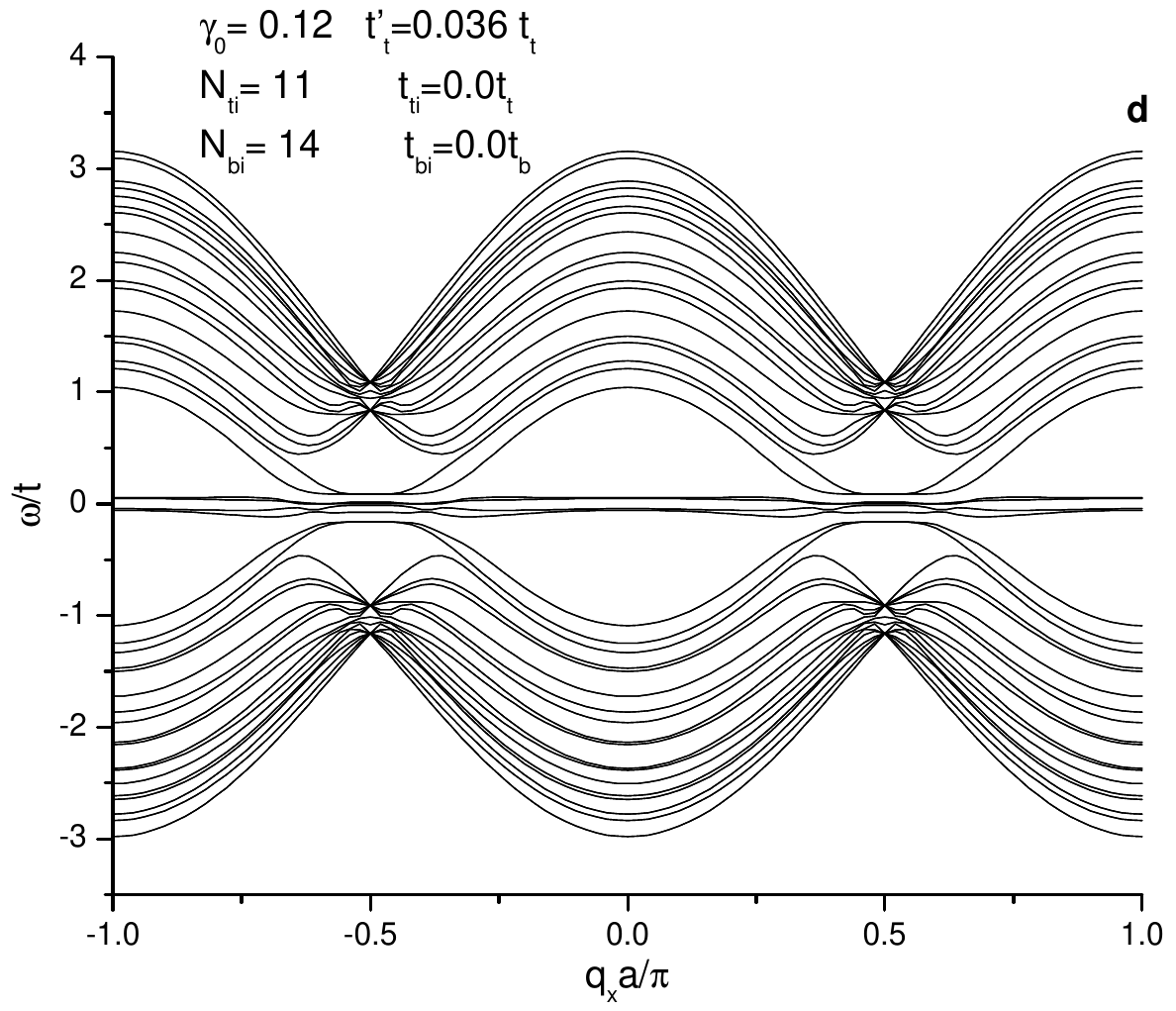}
\end{tabular}
  \caption{The tunable dispersion relations of AA-BLG nanoribbons with zigzag edge and width $N=20$. (a) $\gamma_0=0.0$ (b) $\gamma_0=0.12$ and $t'=0.036t$
  (c)$\gamma_0=0.12$, $t'=0.036t$, $N_{ti} =11$, and $t_{ti}=0.0t_{t}$ (d)$\gamma_0=0.12$, $t'=0.036t$, $N_{ti} =11$, $t_{ti}=0.0t_{t}$, $N_{bi} =14$, and $t_{bi}=0.0t_{b}$.}\label{zigzag207}
\end{figure}

\begin{figure}[hp]
  \centering
  \begin{tabular}{cc}
\includegraphics[scale=.6]{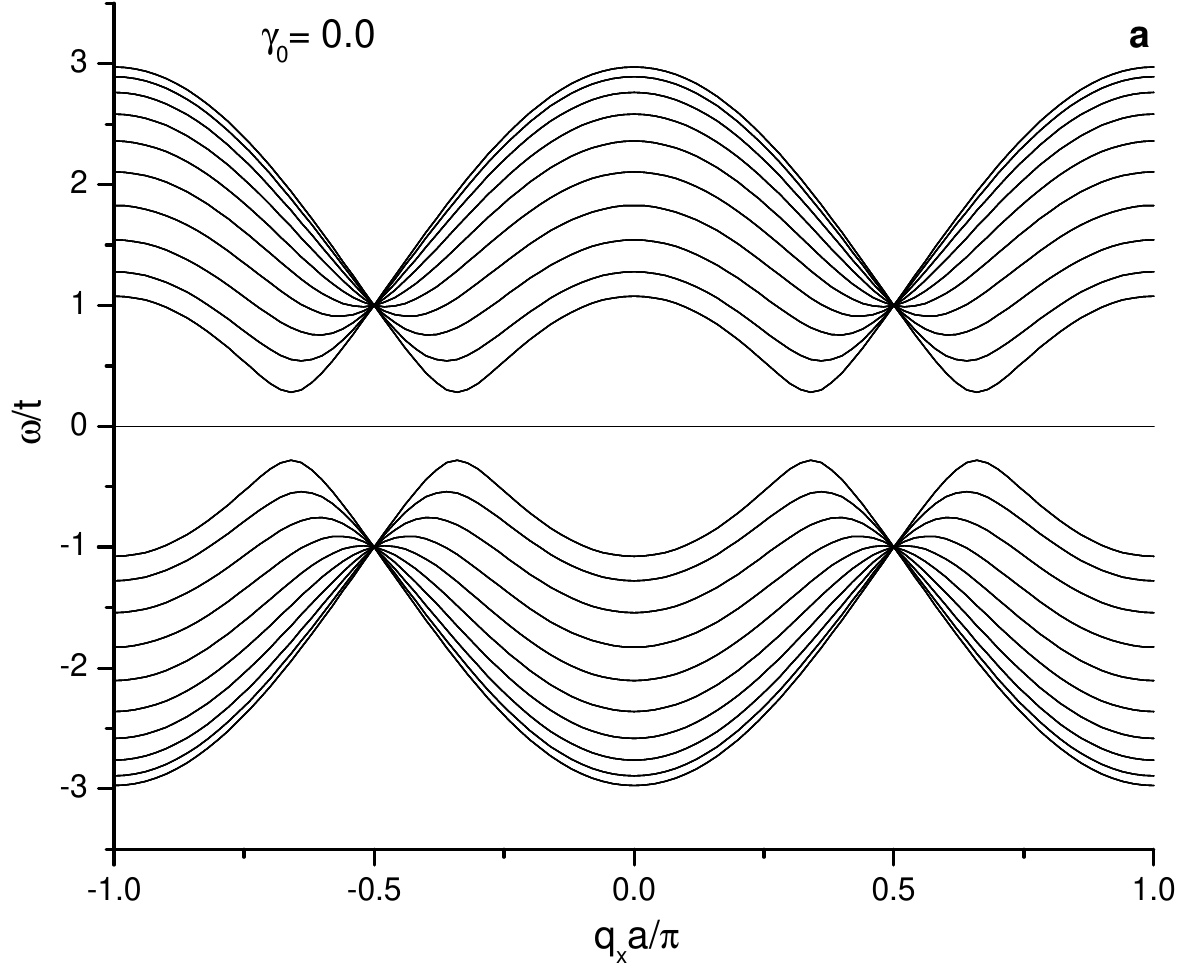} &\includegraphics[scale=.6]{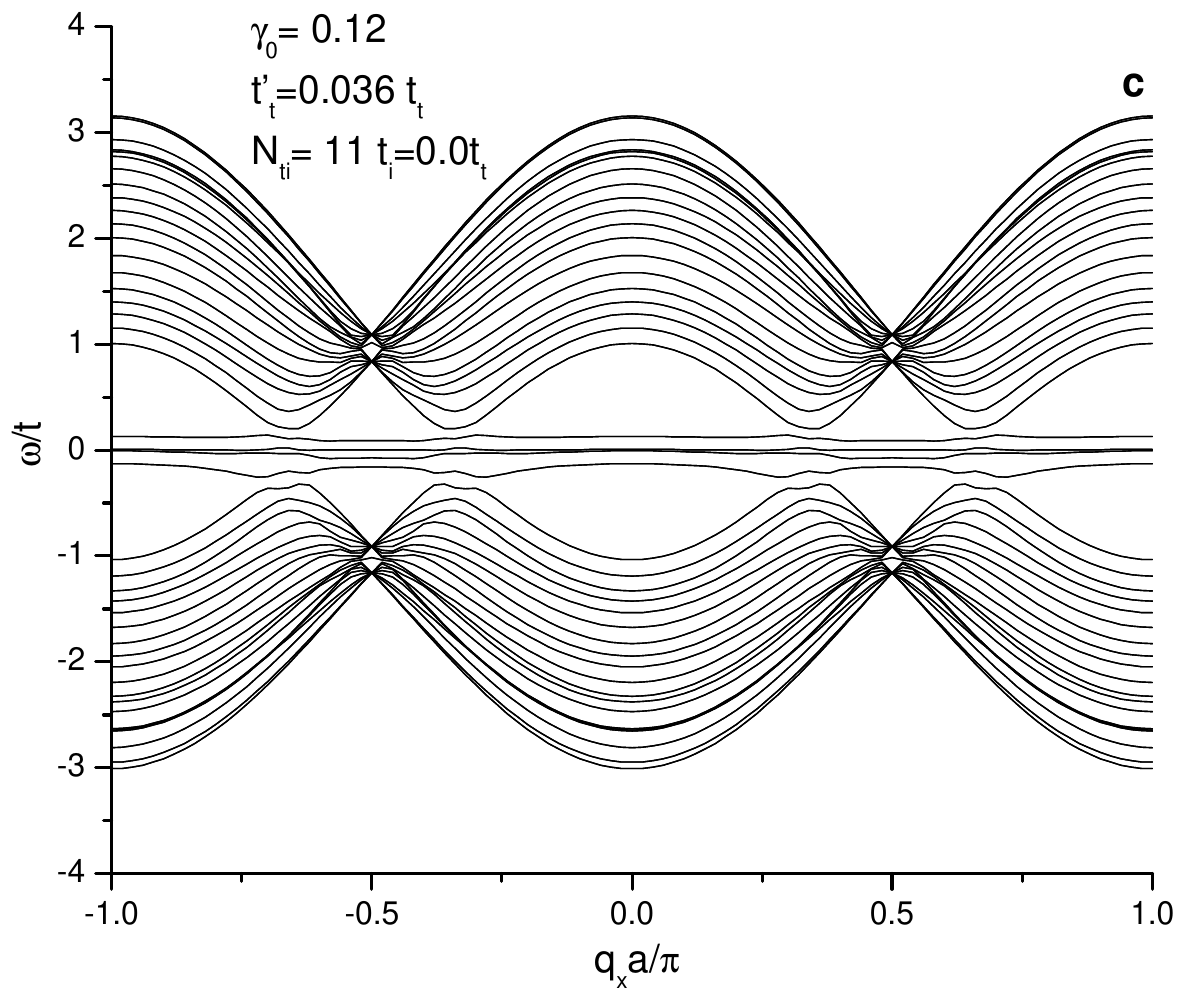}\\
\includegraphics[scale=.6]{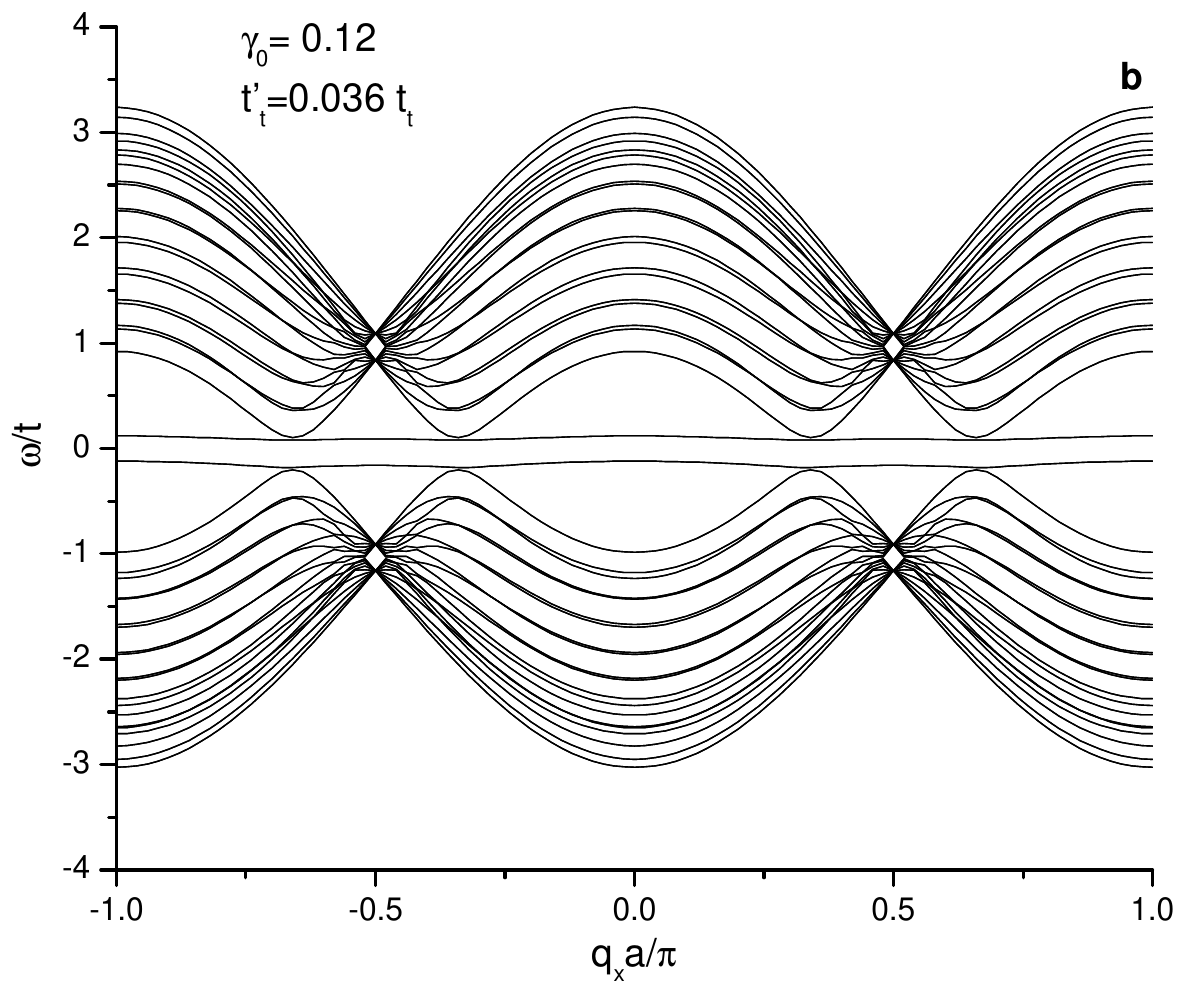}&\includegraphics[scale=.6]{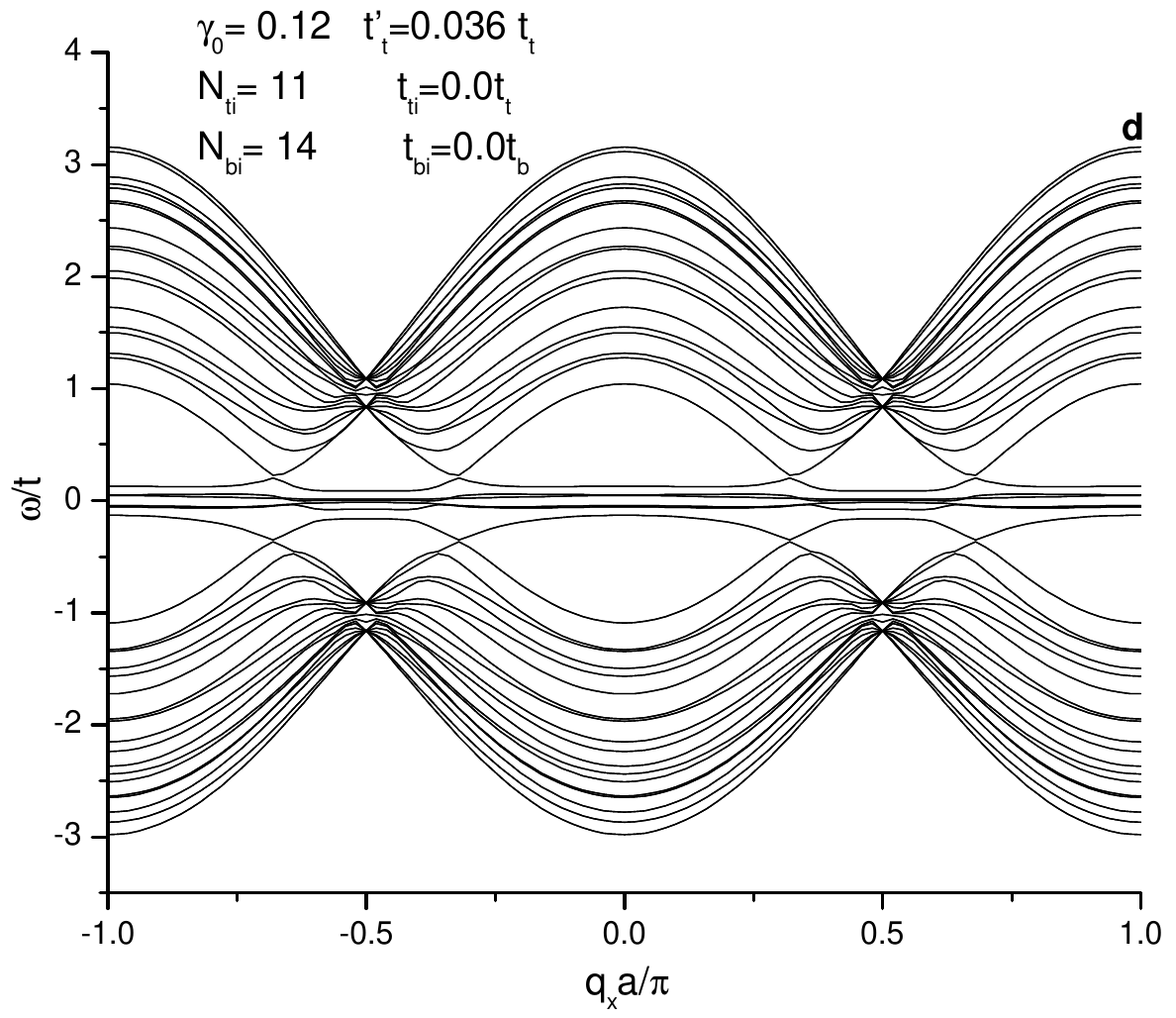}
\end{tabular}
  \caption{The tunable dispersion relations of AA-BLG nanoribbons with zigzag edge and width $N=21$. (a) $\gamma_0=0.0$  (b) $\gamma_0=0.12$ and $t'=0.036t$
  (c)$\gamma_0=0.12$, $t'=0.036t$, $N_{ti} =11$, and $t_{ti}=0.0t_{t}$ (d)$\gamma_0=0.12$, $t'=0.036t$, $N_{ti} =11$, $t_{ti}=0.0t_{t}$, $N_{bi} =14$, and $t_{bi}=0.0t_{b}$.}\label{ert}
\end{figure}

\begin{figure}[h]
  \centering
  \begin{tabular}{cc}
\includegraphics[scale=.6]{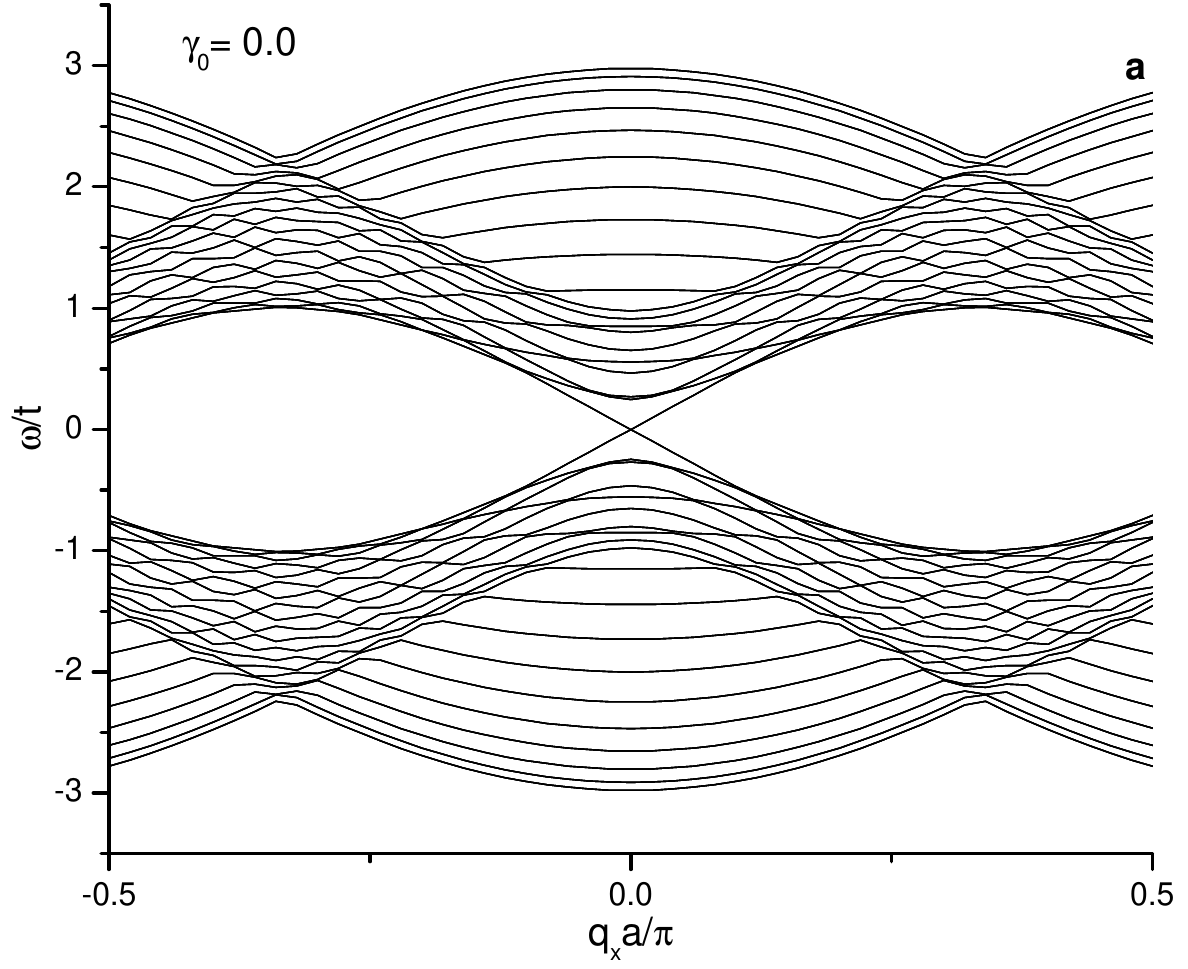}& \includegraphics[scale=.6]{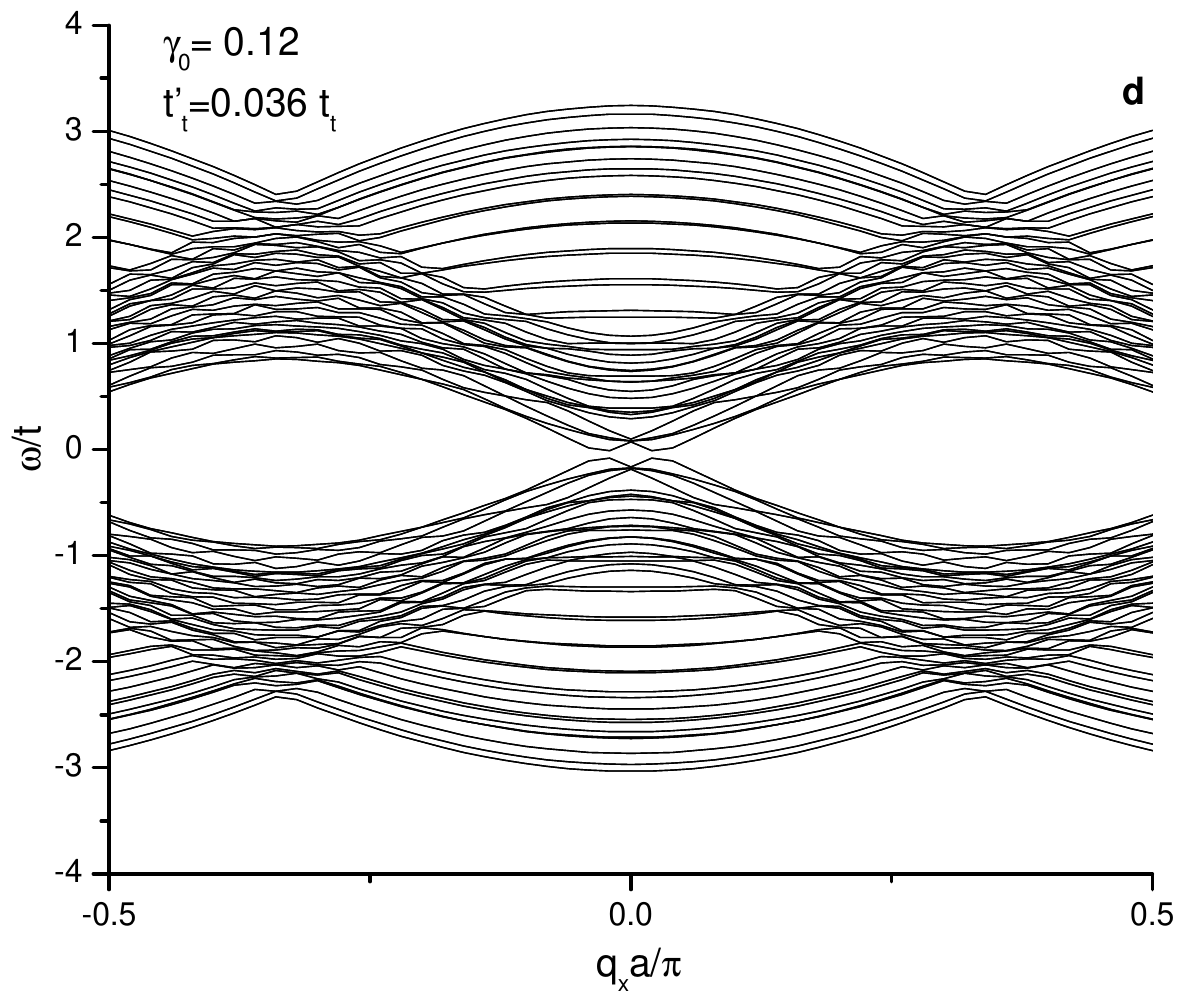}
\end{tabular}
 \caption{The tunable dispersion relations of AA-BLG nanoribbons with armchair edge and width $N=20$. (a) $\gamma_0=0.0$ (b) $\gamma_0=0.12$ and $t'=0.036t$.}\label{armchair207}
\end{figure}

\begin{figure}[h]
  \centering
  \begin{tabular}{cc}
\includegraphics[scale=.6]{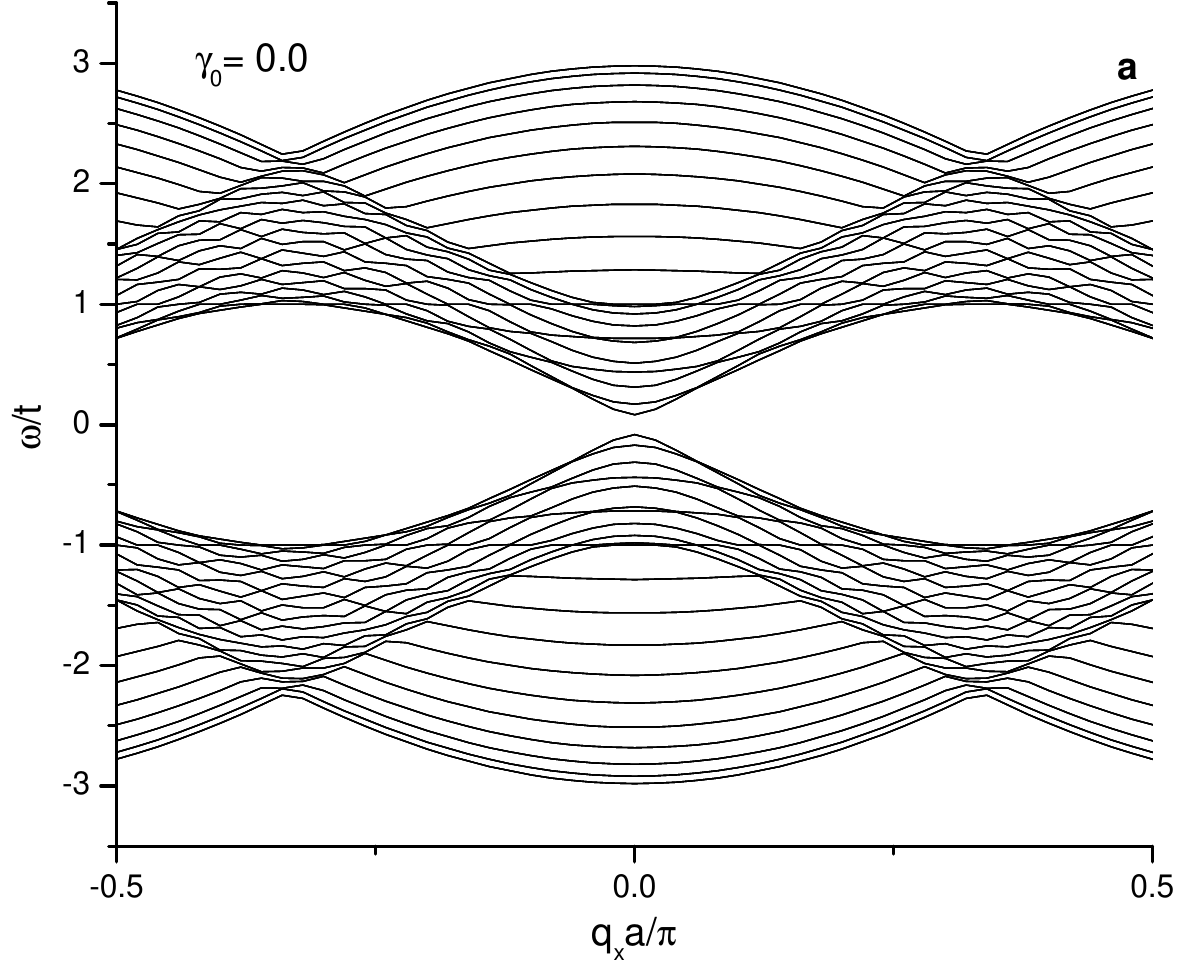}& \includegraphics[scale=.6]{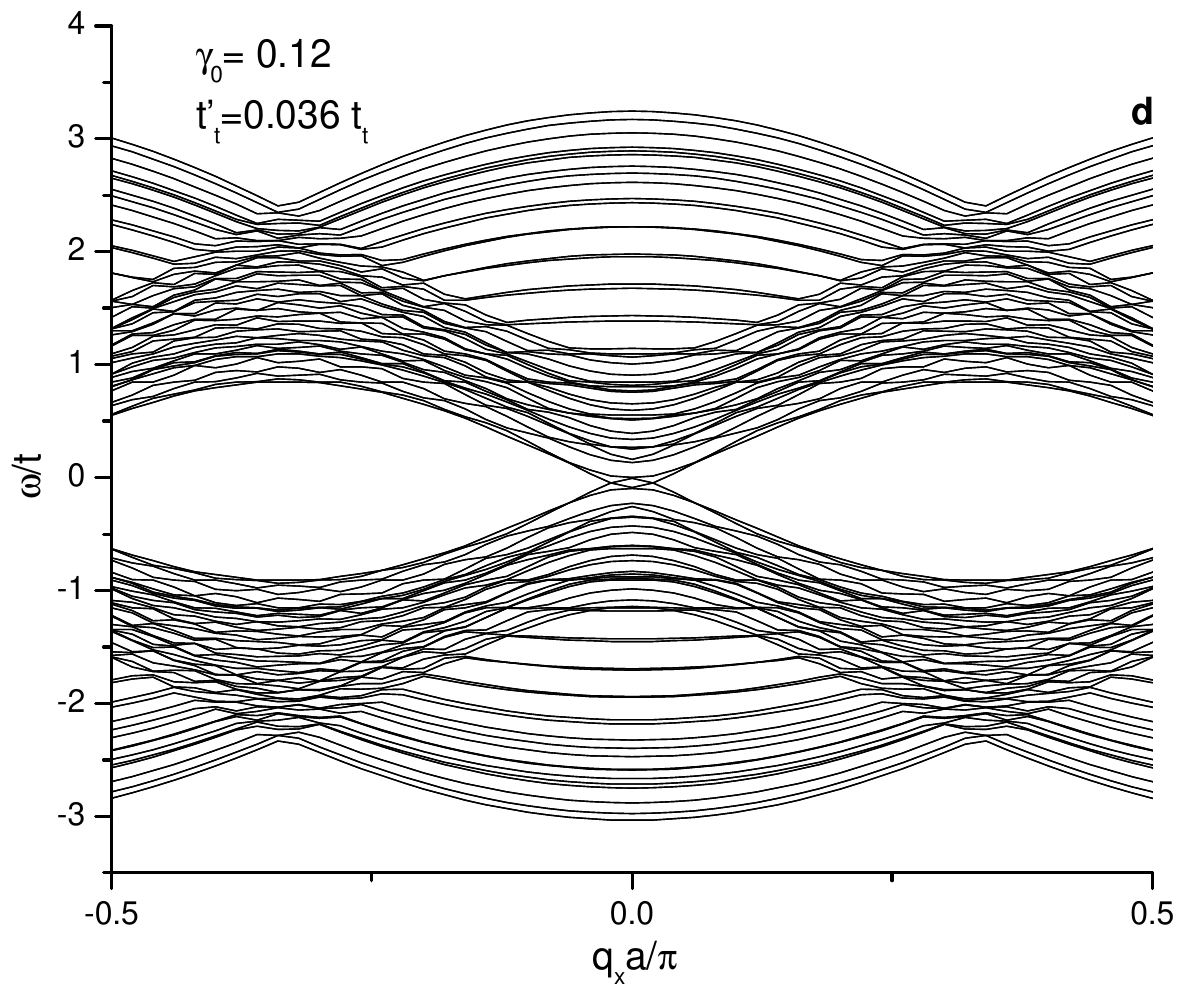}
\end{tabular}
\caption{The tunable dispersion relations of AA-BLG nanoribbons with armchair edge and width $N=21$. (a) $\gamma_0=0.0$ (b) $\gamma_0=0.12$ and $t'=0.036t$.}\label{armchair217}
\end{figure}

\begin{figure}[h]
  \centering
  \begin{tabular}{cc}
\includegraphics[scale=.6]{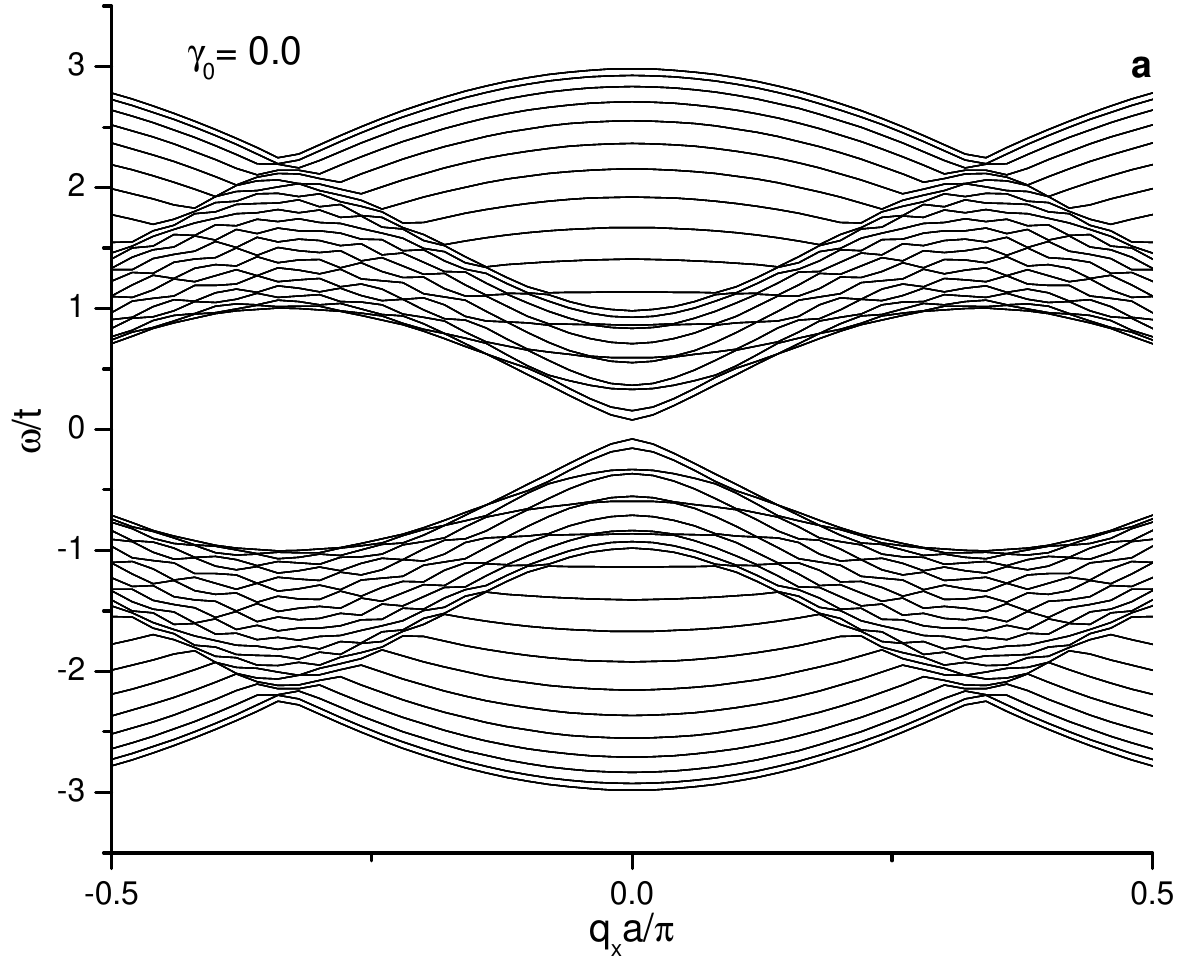}& \includegraphics[scale=.6]{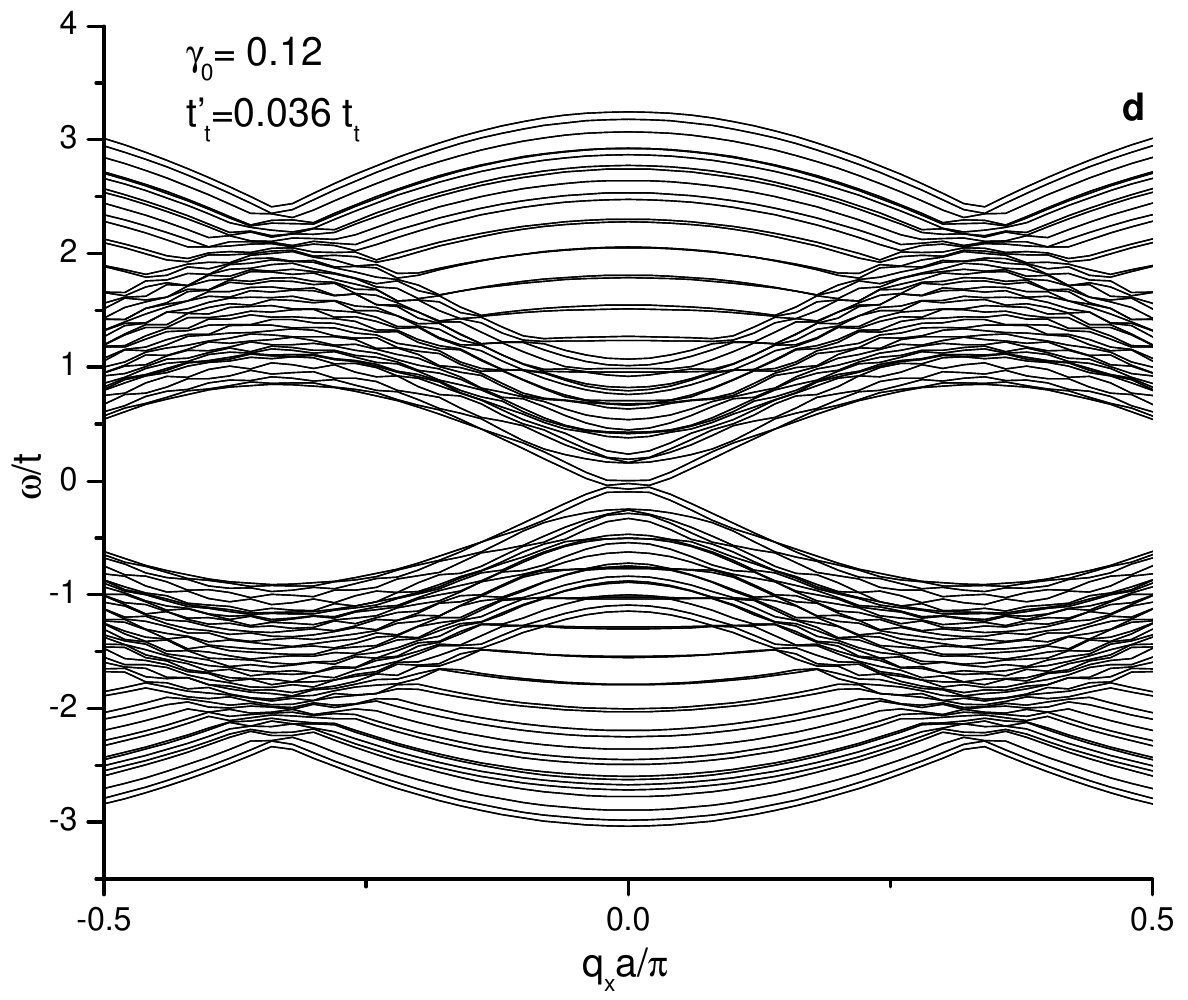}
\end{tabular}
  \caption{The tunable dispersion relations of AA-BLG nanoribbons with armchair edge and width $N=22$. (a) $\gamma_0=0.0$ (b) $\gamma_0=0.12$ and $t'=0.036t$.}\label{armchair227}
\end{figure}

\begin{figure}[h]
  \centering
  \begin{tabular}{cc}
\includegraphics[scale=.6]{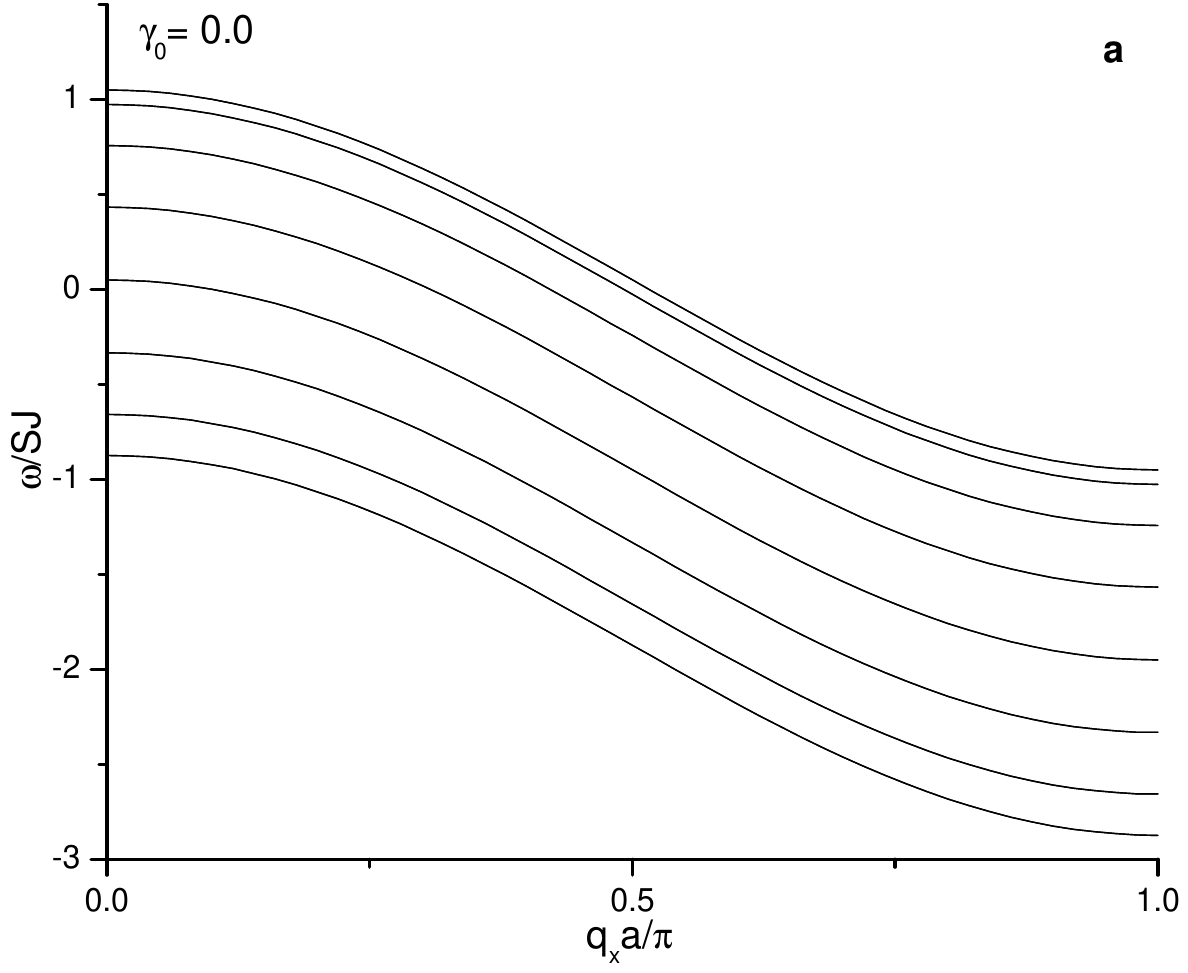}&\includegraphics[scale=.6]{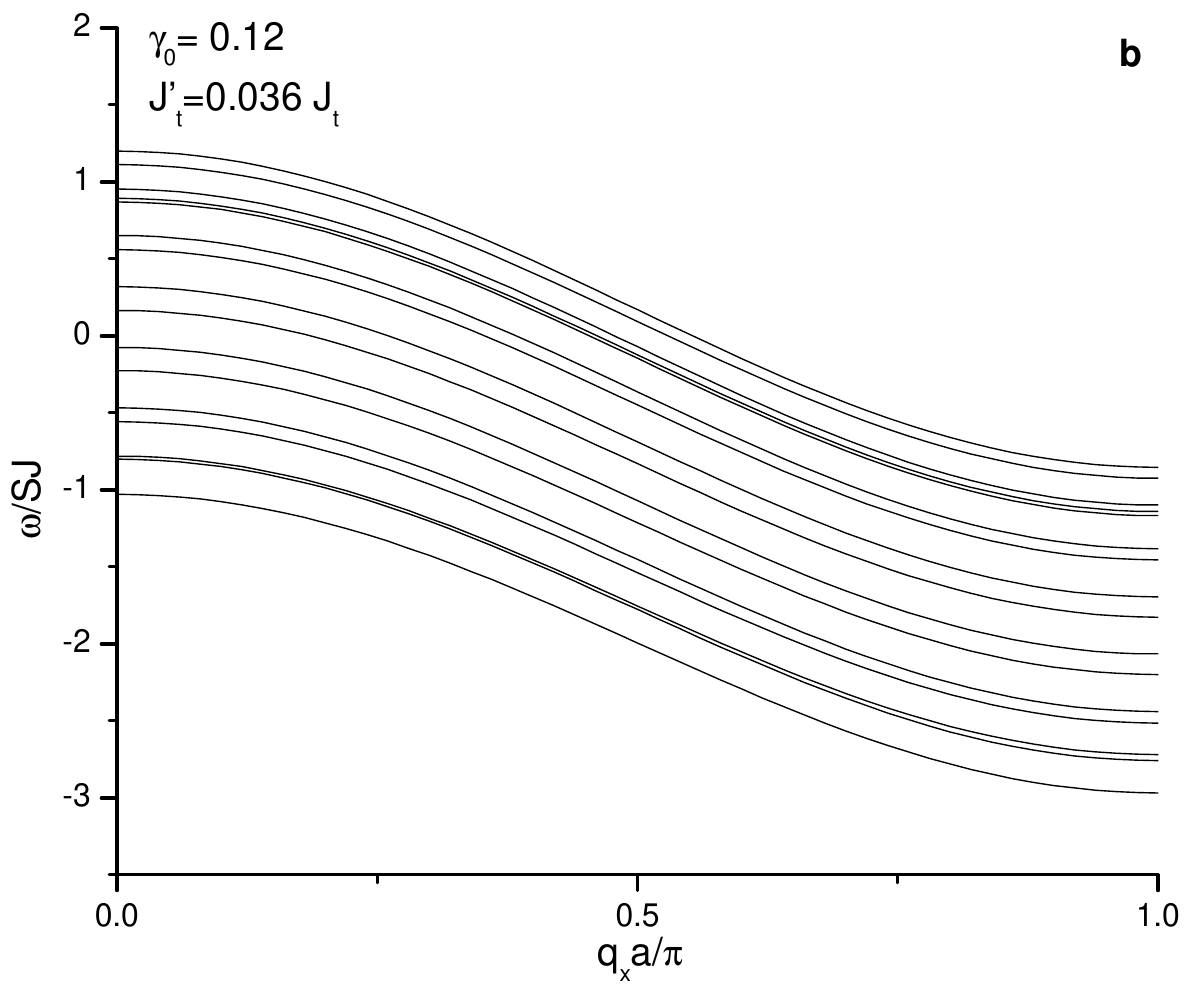}
\end{tabular}
  \caption{For comparison, the tunable dispersion relations of 2D square lattice magnetic stripes with width $N=8$. (a) $\gamma_0=0.0$ (b) $\gamma_0=0.12$ and $t'=0.036t$.}\label{square7}
\end{figure}

In this section the results for the 2D bilayer systems are presented so that
a comparison can readily be made between the different structures (zigzag or
armchair) and different interaction parameters (arising from the interlayer
coupling strength, the range of in-plane hopping, and/or the presence of lines of impurities). These may be
useful for controlling and tuning the mode properties in these ribbon
structures, taking zigzag and armchair edged AA-BLG nanoribbons with all
relevant cases for the width factor. As explained before, these consist of
$N$ even or odd for zigzag structures and $N$ equal to $2i$, $2i+1$, or
$2i+2$ for armchair structures. A brief comparison will also be made with the analogous
case of 2D magnetic square lattice stripes.

Figures \ref{zigzag207}-\ref{ert} show some of the results obtained for the
mode frequencies plotted versus the dimensionless wave vector, taking the above
types of bilayer systems when the ribbon width is even ($N$ = 20) and odd
($N$ = 21), respectively, for the zigzag case. In each figure, panel (a)
shows the dispersion relations of the bilayer systems when there is NN
hopping within the layers and zero inter-layer NN coupling energy
$\gamma_0=0.0$. Consequently, we obtain just the individual layer dispersion
relations, but each mode is now doubly degenerate. Then panel (b) in each
figure shows the effect of introducing an asymmetry between the same coupled
bilayer systems by choosing a nonzero value $\gamma_0=0.12$ and including NNN
hopping with $t'=0.036t$ in the top layer only. This results in an increased
shift in frequency between the two layer modes and those of the single-layer
case in all systems. We note that the wave-vector behavior of the modes near
zero frequency is different in the odd and even $N$ cases. Next, panel (c) in
each of Figures \ref{zigzag207} and \ref{ert} shows the effect of introducing
a line of impurities at row 11 in the top layer only and with impurities
hopping set equal to $t_{ti}=0.0t_{t}$. This causes the introduction of
extended flat localized impurities states at the Fermi level in the
dispersion relations. Finally panel (d) shows the effect of introducing a
second line of impurities, this time in the bottom layer, but now in a
different row (chosen as row 14) and with impurities hopping equal to
$t_{bi}=0.0t_{b}$. This seems to result in an increased tendency towards
degeneracy in the dispersion relations, compared with panel (c), for these
zigzag-edge bilayer systems.

Next, in Figures \ref{armchair207}-\ref{armchair227}, we show some of the
analogous results obtained in the case of studying armchair-edge bilayer
systems, taking width $N$ equal to 20, 21, and 22, respectively. Panels (a)
and (b) in each of these cases refer to similar situations as in the previous
examples (i.e., the limit of uncoupled layers and the asymmetric case of
coupled layers with NNN hopping in one of the layers). The interlayer
coupling effects seem to be rather more significant for the case of $N = 20$
compared with 21 or 22. We have also studied the effects of one or two lines
of impurities. For brevity the results are not presented here, but they are broadly analogous to the zigzag case.

Finally, for comparison purposes, we show in Figure \ref{square7} some results to illustrate interlayer coupling in bilayer magnetic
stripes with a square lattice structure, where the wavevector dependence of the SW modes is significantly different.

\section{Discussion and Conclusions}
In this work the  AA-stacking bilayer graphene nanoribbons were used as an
example of coupled bilayer systems which could be studied using the tight
binding model. The tight binding model calculations for the AA-BLG
nanoribbons show that the bilayer systems can be conveniently analyzed
through an extension of the matrix direct diagonalization method used in \cite{Ahmed4,ahmed5}. This is done by forming two block diagonal matrices with each
block diagonal matrix representing the hopping terms within each individual layer. The
effects of edges, NNN hopping and impurities of a single layer are introduced
numerically as well as new terms for the interlayer coupling.

The bilayer results in this work can also be generalized to multilayered systems in general,
provided they are formed by direct stacking of identical 2D layers with the same lattices and geometries as considered here.

The obtained dispersion relations for zigzag and armchair AA-BLG nanoribbons
and magnetic square bilayer stripes show that the bilayer systems offer more
flexibility with regards to the possibility of tuning their properties by
changing parameters such as the interlayer hopping strength by
changing the interlayer distance, adding different impurities configurations
in individual layers, and changing the range of the interaction in individual
layers. Also, the results show that the
sensitivity of the bilayer systems to these parameters is strongly dependent on their
lattice structure.

\begin{acknowledgments}
This research has been supported by the Egyptian Ministry of Higher Education
and Scientific Research (MZA).
\end{acknowledgments}

\bibliography{xbib2}% Produces the bibliography via BibTeX.

\end{document}